\documentclass[11pt]{article}
\usepackage{ifthen}
\usepackage{times,mathptm,color}
\usepackage{graphicx}
\usepackage{amsthm, amsmath, amssymb, enumerate, hyperref}

\newcommand{\change}[1]{}

\theoremstyle{plain}
\newtheorem{theorem}{Theorem}[section]
\newtheorem{lemma}[theorem]{Lemma}
\newtheorem{corollary}[theorem]{Corollary}
\newtheorem{proposition}[theorem]{Proposition}

\theoremstyle{definition}
\newtheorem{definition}[theorem]{Definition}
\newtheorem{example}{Example}

\theoremstyle{remark}

\def\squareforqed{\hbox{\rlap{$\sqcap$}$\sqcup$}}
\def\qed{\ifmmode\squareforqed\else{\unskip\nobreak\hfil
\penalty50\hskip1em\null\nobreak\hfil\squareforqed
\parfillskip=0pt\finalhyphendemerits=0\endgraf}\fi}

\newenvironment{proofof}[1]{\begin{trivlist}%
\item[]{\flushleft\em Proof of #1. }}
{\qed\end{trivlist}}

\newcommand{\comments}[1]{}
\newcommand{\maxDegree}{\Delta}
\newcommand{\degree}{d}
\newcommand{\tw}{\mathrm{tw}}
\newcommand{\cc}{\mathrm{cc}}

\newcommand{\Hspace}{\mathcal{H}} \newcommand{\linear}{\mathbf{L}}

\newcommand{\defeq}{\stackrel{\mathrm{def}}{=}}

\newcommand{\sepAuthor}{.5in}
\newcommand{\sepAbstract}{.5in}
\newcommand{\skipKeywords}{30pt}
\newcommand{\sepTitle}{2ex}

\long\def\mytitlepage#1#2#3#4{
        \thispagestyle{empty}
    \vspace*{\sepTitle}
        \begin{center}
        {\Large\bf #1}

        \vspace{\sepAuthor}
        #2\\
        \medskip

        \vspace{\sepAbstract}
        {\Large Abstract}
        \end{center}

        \noindent{#3}
        \vskip\skipKeywords

        \noindent{#4}
        \clearpage
        }
\begin{document}
\mytitlepage{
Simulating quantum computation by contracting tensor networks
}
{{\large Igor L. Markov\footnote{Supported in part
by NSF 0208959, the DARPA QuIST program and the Air Force Research Laboratory.}\ \  and\ \  Yaoyun Shi\footnote{Supported in part
by NSF 0323555, 0347078 and 0622033.}  }\\
\vspace{1ex}
Department of Electrical Engineering and Computer Science\\
The University of Michigan\\
2260 Hayward Street\\
Ann Arbor, MI 48109-2121, USA\\
E-mail: \{imarkov$|$shiyy\}@eecs.umich.edu}
{
The treewidth of a graph is a useful combinatorial
measure of how close the graph is to a tree. 
We prove that a quantum circuit with $T$ gates whose 
underlying graph has treewidth $d$ can be simulated 
deterministically in $T^{O(1)}\exp[O(d)]$ time, which, 
in particular, is polynomial in $T$ if $d=O(\log T)$.
Among many implications, we show efficient simulations 
for log-depth circuits whose gates apply to 
nearby qubits only, a natural constraint satisfied by
most physical implementations. We also 
show that {\em one-way quantum computation}
of Raussendorf and Briegel ({\em Physical Review Letters}, 86:5188--5191, 2001),
a universal quantum computation scheme with promising physical implementations,
can be efficiently simulated by a randomized algorithm if its quantum resource
is derived from a small-treewidth graph.
}{{\bf Keywords}: Quantum computation, computational complexity,
treewidth, tensor network, classical simulation, one-way quantum computation.}

\section{Introduction} 
The recent interest in quantum circuits is motivated by several complementary
considerations. Quantum information processing is rapidly becoming a reality as
it allows manipulating matter at unprecedented scale. Such manipulations may
create particular entangled states or implement specific quantum evolutions ---
they find uses in atomic clocks, ultra-precise metrology, high-resolution
lithography, optical communication, etc. On the other hand, engineers
traditionally simulate new designs before implementing them. Such simulation
may identify subtle design flaws and save both costs and effort. It typically
uses well-understood host hardware, e.g., one can simulate a quantum circuit
on a commonly-used conventional computer.

More ambitiously, quantum circuits compete with conventional computing
and communication. Quantum-mechanical effects may potentially lead to
computational speed-ups, more secure or more efficient communication, better
keeping of secrets, etc. To this end, one seeks new circuits and algorithms
with revolutionary behavior as in Shor's work on number-factoring, or provable
limits on possible behaviors. While proving abstract limitations on the success
of unknown algorithms appears more difficult, a common line of reasoning for
such results is based on simulation. For example, if the behavior of a quantum
circuit can be faithfully simulated on a conventional computer, then the
possible speed-up achieved by the quantum circuit is limited by the cost of
simulation. Thus, aside from sanity-checking new designs for quantum
information-processing hardware, more efficient simulation can lead to sharper
bounds on all possible algorithms.


Since the outcome of a quantum computation is probabilistic, we shall clarify
our notion of simulation. By a randomized simulation, we mean a classical
randomized algorithm whose output distribution on an input is identical
to that of the simulated quantum computation. By a deterministic simulation,
we mean a classical deterministic algorithm which, on a given pair of input
$x$ and output $y$ of the quantum computation, outputs the probability that
$y$ is observed at the end of the quantum computation on $x$.

To simulate a quantum circuit, one may use a na{\"\i}ve brute-force
calculation of quantum amplitudes that has exponential overhead.
Achieving significantly smaller overhead in the generic case appears
hopeless --- in fact, this observation lead Feynman to suggest
that quantum computers may outperform conventional ones in some
tasks.  Therefore, only certain restricted classes of quantum
circuits were studied in existing literature on simulation.

Classes of quantum circuits that admit efficient simulation are often
distinguished by a restricted ``gate library'', but do not impose additional
restrictions on how gates are interconnected or sequenced. A case in point is
the seminal Gottesman-Knill Theorem~\cite{Gottesman:1998:simulate} and its
recent improvement by Aaronson and Gottesman~\cite{AaronsonG04}. These results
apply only to circuits with stabilizer gates --- Controlled-NOT, Hadamard,
Phase, and single-qubit measurements in the so called Clifford group.
\comments{
While stabilizer gates are not universal,
the main appeal of this work is that it directly applies to error-correcting
circuits and helps estimating thresholds for error-tolerance required for
scalability.}
Another example is given by {\em match gates} defined and studied
by Valiant~\cite{Valiant:2002:simulate}, and extended by Terhal and DiVincenzo
\cite{Terhal:2002:simulate}.

A different way to impose a restriction on a class of quantum circuits is to
limit the amount of entanglement in intermediate states.  Jozsa and Linden
\cite{JL02a}, as well as Vidal~\cite{Vidal03} demonstrate efficient classical
simulation of such circuits and conclude that achieving
quantum speed-ups requires more than a bounded amount of entanglement.

In this work we pursue a different approach to efficient simulation and allow
the use of arbitrary gates.
More specifically, we assume a general quantum circuit model in which
a gate is a general quantum operation (so called {\em physically realizable operators})
on a constant number of qubits. This model, proposed and studied
by Aharonov, Kitaev and Nisan~\cite{AKN98a}, generalizes the standard quantum circuit
model, defined by Yao~\cite{Yao93},
where each gate is unitary and measurements are applied at the end of the computation.
We also assume that (i) the computation
starts with a fixed unentangled state in the computational basis,
and (ii) at the end each qubit is either measured or {\em traced-out}.

Our simulation builds upon the framework of {\em tensor network contraction}.
Being a direct generalization of matrices,
tensors capture a wide range of linear phenomena including vectors,
operators, multi-linear forms, etc. They facilitate convenient and
fundamental mathematical tools in many branches of physics such as
fluid and solid mechanics, and general relativity~\cite{Joshi95}.
More recently, several methods have been developed to simulate
quantum evolution by contracting variants of tensor networks,
under the names of {\em Matrix Product States (MPS)}, 
{\em Projected Entangled Pairs States (PEPS)}, etc \cite{Vidal03, Vidal04,
Verstraete04, Zwolak04, Verstraete04, VerstraeteG04, Porras05}.
Under this framework, a quantum circuit is regarded
as a network of tensors. The
simulation contracts edges one by one and performs the 
convolution of the corresponding tensors, 
until there is only one vertex left. Having degree 0,
this vertex must be labeled by a single number, which gives the
final measurement probability sought by simulation. In contrast with other
simulation techniques, we do not necessarily simulate individual
gates in their original order --- in fact, a given gate may even be
simulated {\em partially} at several stages of the simulation.

While tensor network contraction has been used in previous work,
little was known about optimal contraction orders.
We prove that the minimal cost of contraction is determined by
the {\em treewidth} $\tw(G_C)$ of the circuit graph $G_C$.
Moreover, existing constructions that approximate optimal
{\em tree-decompositions} (e.g.~\cite{RSX}) 
produce near-optimal contraction sequences.
We shall define the concepts of treewidth and tree decompositions
in Section 2. Intuitively, the smaller a graph's treewidth is,
the closer it is to a tree, and a tree decomposition is a drawing
of the graph to make it look like a tree as much as possible.
Our result allows us to
leverage the extensive graph-theoretical literature dealing with the properties
and computation of treewidth.

\begin{theorem}\label{res:main}
Let $C$ be a quantum circuit with $T$ gates and whose underlying
circuit graph is $G_C$. Then
$C$ can be simulated deterministically 
in time $T^{O(1)}\exp[O(\tw(G_C))]$.
\end{theorem}

A rigorous restatement of the above theorem is Theorem~\ref{res:simuCCTW}.
By this theorem, given a function computable in polynomial time by a
quantum algorithm but not classically, any polynomial-size quantum
circuit computing the function must have super-logarithmic
treewidth.
\comments{
The {\em simulation} in the above informal statement is deterministic and
produces more information than an actual quantum circuit would: given an
arbitrary measurement on an arbitrary subset of the qubits in the computational
basis, and an outcome of the measurement, it gives the exact probability of
observing this outcome.
}

The following corollary is an immediate consequence.
\begin{corollary}\label{res:polysimu}
Any polynomial-size quantum circuit of a logarithmic treewidth can
be simulated deterministically in polynomial time.
\end{corollary}
Quantum formulas defined and studied by Yao
\cite{Yao93} are quantum circuits whose underlying graphs are
trees. Roychowdhury and Vatan~\cite{RoychowdhuryV} showed that quantum formulas can be efficiently
simulated deterministically. Since every quantum formula has treewidth 1,
Corollary~\ref{res:polysimu} gives an alternative efficient simulation.

  Our focus on the {\em topology} of the quantum circuit allows us to
  accommodate arbitrary gates, as long as their qubit-width (number of inputs)
  is limited by a constant. In particular, Corollary \ref{res:polysimu} implies
  efficient simulation of some circuits that create the maximum amount of entanglement
  in a partition of the qubits, e.g., a layer of two-qubit gates. Therefore, our
  results are not implied by previously published techniques.

We now articulate some implications of our main result to classes of
quantum circuits, in terms of properties of their underlying graphs.
The following two classes of graphs are well-studied, and their
treewidths are known. The class of {\em series parallel graphs}
arises in electric circuits, and such circuits have treewidth
$\le2$. Planar graphs $G$ with $n$ vertices are known to have
treewidth $\tw(G)=O(\sqrt{|V(G)|})$~\cite{AlberB+02}.

\begin{corollary}\label{res:paral}
Any polynomial size parallel serial quantum circuit can
be simulated deterministically in polynomial time.
\end{corollary}

\begin{corollary}\label{res:planar}
A size $T$ planar quantum circuit can be simulated deterministically
in $\exp[O(\sqrt{T})]$ time.
\end{corollary}

Another corollary deals with a topological restriction
representative of many physical realizations of quantum circuits.
Let $q\ge 1$ be an integer. A circuit is said to be 
$q$-local-interacting if under a linear ordering of its qubits, each gate 
acts only on qubits that are at most $q$ distance apart.
A circuit is said to be local-interacting if it is $q$-local interacting
with a constant $q$ independent of the circuit size. 
Such {\em local-interaction} circuits generalize
the restriction of qubit couplings to nearest-neighbor qubits (e.g.,
in a spin-chain) commonly appearing in proposals for building
quantum computers, where qubits may be stationary and cannot be
coupled arbitrarily. To this end, we observe that the treewidth of
any local-interaction circuit of logarithmic depth is at most
logarithmic.

\begin{corollary}\label{res:near}
Let $C$ be a quantum circuit of size $T$ and depth $D$, and is
$q$-local-interacting.
Then $C$ can be simulated deterministically in
$T^{O(1)}\exp[O(qD)]$ time. In particular, if $C$ is a
polynomial-size local-interacting circuit with a logarithmic depth,
then it can be simulated deterministically in polynomial time.
\end{corollary}

Yet another important application of our approach is to the
simulation of {\em one-way quantum computation}. In two influential
papers~\cite{BR, RB}, Briegel and Raussendorf introduced the concept
of {\em graph states} --- quantum states derived from graphs, ---
and show that an arbitrary quantum circuit can be simulated by {\em
adaptive}, {\em single-qubit measurements} on the {\em graph state}
derived from the grid graph.
Note that the graph state for a one-way quantum computation
does not depend on the quantum circuit to be simulated (except
that its size should be large enough) and that for most
physical implementations single-qubit measurements are much
easier to implement than multi-qubit operations.
Hence it is conceivable that graph states would be manufactured
by a technologically more advanced party, then used by other parties
with lesser quantum-computational power in order to facilitate
universal quantum computing.
This makes one-way quantum computation an attractive scheme 
for physical implementations of universal quantum computation.
An experimental demonstration of one-way
quantum computation appeared in a recent Nature
article~\cite{Walther05}.

A natural question about one-way computation is to characterize the
class of graphs whose graph states are universal for quantum
computation. We call a family of quantum states $\phi=\{|\phi_1\rangle, |\phi_2\rangle, \cdots,
|\phi_n\rangle, \cdots\}$ {\em universal for one-way quantum computation}
if (a) the number of qubits in $|\phi_n\rangle$ is bounded by a fixed polynomial
in $n$; (b) any quantum circuit of size $n$ can be simulated by a one-way
quantum computation on $|\phi_n\rangle$.
On the other hand, $\phi$ is said to be efficiently simulatable if
any one-way quantum computation on $|\phi_n\rangle$ can be efficiently simulated
classically for all sufficiently large $n$. Note that the class of universal
families and that of efficiently simulatable families are disjoint if and only if efficient quantum
computation is indeed strictly more powerful than efficient classical computation.
We show that it is necessary for graphs to have high treewidth so that the corresponding
graph states are not efficiently simulatable.

\begin{theorem}\label{res:oneway}
Let $G$ be a simple undirected graph. Then a one-way quantum
computation on the respective graph state can be simulated by a
randomized algorithm in time $|V(G)|^{O(1)}\exp[O(\tw(G))]$.
\end{theorem}

Our simulation can be made deterministic with a better upper bound on time 
complexity if the one-way computation satisfies additional constraints,
such as those in \cite{RB}. We shall elaborate on this improvement
in Section~\ref{sec:oneway}.

 An important limitation of our techniques is that
a circuit family with sufficiently fast-growing treewidth may 
require super-polynomial resources for simulation. In particular, 
this seems to be the case with known circuits for modular exponentiation.
Therefore, there is little hope to efficiently
simulate number-factoring algorithms using tree decompositions. 
As an extreme example to illustrate the limitation of our technique, we give a depth-$4$ circuit --- 
including the final measurement as the 4th
layer --- that has large treewidth.

\begin{theorem}\label{res:depth4}
There exists a depth-4
quantum circuit on $n$ qubits using only one- and two-qubit gates
such that its treewidth is $\Omega(n)$.
\end{theorem}

Note that a circuit satisfying the assumption in the above theorem
must have $O(n)$ size.
Our construction is based on expander graphs, 
whose treewidth must be linear
 in the number of vertices (Lemma~\ref{res:expander}).

This finding is consistent with the obstacles to efficient
simulation that are evident in the results of Terhal and DiVincenzo
\cite{Terhal:2002:constant}, later extended by Fenner et al.
\cite{FennerGHZ04}. In contrast, we are able to efficiently simulate
any depth-$3$ circuit {\em deterministically} while the simulation
in~\cite{Terhal:2002:constant} is probabilistic.

\begin{theorem}\label{res:depth3}
Assuming that only one- and two-qubit gates are allowed, any
polynomial-size depth-$3$ quantum circuit can be simulated
deterministically in polynomial time.
\end{theorem}

Our simulation algorithm is related to algorithms for other tasks in
that its runtime depends on the treewidth of a graph derived from
the input. Bodlaender wrote an excellent survey ~\cite{Bodlaender06} 
on this subject. Particularly relevant are algorithms based on
``vertex eliminations'', e.g., the {\em Bucket Elimination}
algorithm for Bayesian Inference~\cite{Dechter99}. 
Another parallel can be made with the work by
Broering and Lokam~\cite{BroeringL03}, which solves \textrm{Circuit-SAT}
in time exponential in the treewidth of the graph of
the given circuit. However, to our best knowledge,
we are the first to relate the treewidth of a quantum circuit
to its classical simulation.

Our results are applicable to the simulation of classical
probabilistic circuits, which can be modeled by matrices, similarly
to quantum circuits. Such simulation has recently gained prominence
in the literature on the reliability of digital logic
\cite{KrishnaswamyVMH05}, and is particularly relevant to
satellite-based and airborne electronics which experience
unpredictable particle strikes at higher rates.

The rest of the paper is organized as follows. After introducing
notation, we describe how quantum circuits and their simulation can
be modeled by tensor networks. The runtime of such simulation
depends on the graph parameter that we call the {\em contraction
complexity}. We then relate the contraction complexity to treewidth,
and apply the simulation to restricted classes of graphs,
and to one-way quantum computation. Finally, we discuss
possible directions for future investigations
with a brief survey on the subsequent development since 
the announcement of our results.


\section{Notation and definitions}
For integer $n\ge 1$, define $[n]\defeq\{1, 2, \ldots, n\}$. An
ordering $\pi$ of an $n$-element set is denoted by $\pi(1)$,
$\pi(2)$, $\ldots$, $\pi(n)$.
Unless otherwise stated, graphs in this paper are undirected and may have multiple edges or
loops. Edges connecting the same pair of vertices are called parallel edges.
If $G$ is a graph, its vertex set is denoted by $V(G)$ and
its edge set by $E(G)$. When it is clear in the context, we use
$V=V(G)$ and $E=E(G)$. The {\em degree} of a vertex $v$, denoted by $d(v)$, is the number
of edges incident to it. In particular, a loop counts as $1$ edge.
The maximum degree of a vertex in $G$ is denoted by $\maxDegree(G)$.

\begin{trivlist}
\item {\bf Treewidth of a graph.} Let $G$ be a graph. A {\em tree
decomposition} of $G$~\cite{RSIII} is a tree $\mathcal{T}$, together
with a function that maps each vertex $w\in V(\mathcal{T})$ to a subset
$B_w\subseteq V(G)$. These subsets $B_w$ are called {\em bags}
(of vertices). In addition, the following conditions must hold.
\begin{enumerate}[(T1)]
\item $\bigcup_{v\in V(\mathcal{T})} B_v = V(G)$, i.e., each vertex 
      must appear in at least one bag.
\item $\forall \ \{ u, v\}\in E(G)$, $\exists w\in V(\mathcal{T})$,
$\{u, v\}\subseteq B_w$, i.e., for each edge,
at least one bag must contain both of its end vertices.
\item $\forall\ u\in V(G)$, the set of vertices $w\in V(\mathcal{T})$
with $u\in B_w$ form a connected subtree, i.e., all bags containing
a given vertex must be connected in $\mathcal{T}$.
\end{enumerate}
The {\em width} of a tree decomposition is defined by $\max_{w\in
V(\mathcal{T})} |B_w|-1$. The {\em treewidth} of $G$ is the minimum
width over its tree decompositions. For example, all trees have treewidth 1
and single cycles of length at least 3 have treewidth 2.
Figure~\ref{fig:treedecomp} shows an example of tree decomposition.
\begin{figure}[tbph]
\centering
\includegraphics[width=5in]{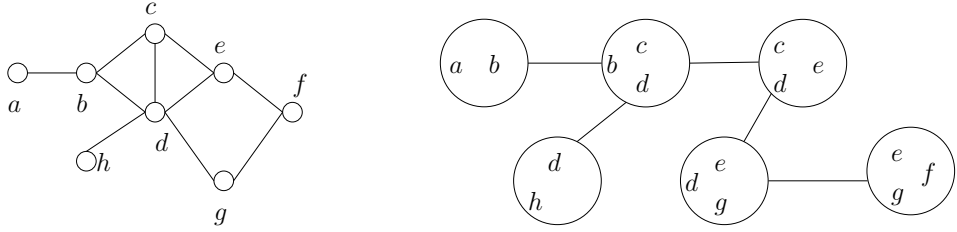}
\caption{A graph and its decomposition of width 2 with 6 bags.}
\label{fig:treedecomp}
\end{figure}
Intuitively, a tree decomposition $\mathcal{T}$
is a way of drawing a graph to look like a tree, which
may require viewing sets of vertices (bags) as single vertices. 
The less a graph looks like a tree, the larger the bags become.
The notion of tree decomposition has been
useful in capturing the complexity of constraint satisfaction
problems, Bayesian networks and other combinatorial phenomena
represented by graphs.
In further writing, we may refer to a vertex in 
$\mathcal{T}$ by its bag when the context is clear. 

Treewidth can be defined in several seemingly unrelated ways, e.g.,
as the minimum $k$ for which a given graph is a {\em partial
$k$-tree}, as the {\em induced width} (also called the {\em
dimension}), or as the {\em elimination width}~\cite{Rose70,
Arnborg85}. An {\em elimination ordering} $\pi$ of a graph $G$ is an
ordering of $V(G)$. The {\em induced width of a vertex} $v\in V(G)$
in the ordering is the number of its neighbors at the time it is
being removed in the following process: start with $\pi(1)$,
add an edge for each pair of its neighbors that were previously not adjacent,
remove $\pi(1)$, then repeat this procedure
with the next vertex in the ordering. The {\em width of $\pi$} is
the maximum induced width of a vertex, and the {\em induced width of
$G$} is the minimum width of an elimination ordering. It is known
that the induced width of a graph is precisely its treewidth
\cite{Arnborg85}.

It follows straightforwardly from the definition of treewidth
that if $G$ is obtained from $G'$ by removing a degree $1$ vertex,
$\tw(G)=\tw(G')$, unless $G'$ has only $1$ edge, in which case
$\tw(G)=0$ and $\tw(G')=1$.
We will also use the following well known and simple fact,
a proof for which is provided in the Appendix.

\begin{proposition}\label{res:insert} Let $G$ be a simple undirected graph, and 
$w$ be a degree $2$ vertex. Then removing $w$ and 
connecting its two adjacent vertices does not change the treewidth.
\end{proposition}

\item {\bf Quantum circuits.}
We review some basic concepts of quantum mechanics and quantum
computation. For a more detailed treatment, we refer the readers
to the book by Nielsen and Chuang~\cite{NielsenC}.

The state space of one qubit is denoted by $\Hspace\defeq \mathbb{C}^2$.
We fix an orthonormal basis for $\Hspace$ and label
the basis vectors with $|0\rangle$ and $|1\rangle$. 
The space of operators on a vector space $V$ is denoted by $\linear(V)$. 
The identity operator on $V$ is denoted by $I_V$, or by $I$ if $V$ is implicit
from the context.
A density operator, or a mixed state, of $n$ qubits
is a positive semi-definite operator $\rho\in\linear(\Hspace^{\otimes n})$
with $\textrm{trace}{\rho}=1$. For a binary string $x=x_1x_2\cdots x_n\in\{0, 1\}^n$, 
let $\rho_x\defeq \bigotimes_{i=1}^n |x_i\rangle\langle x_i|$
be the density operator of the state $|x\rangle\defeq\otimes_{i=1}^n|x_i\rangle$.

In this paper, a quantum gate with $a$ input qubits and $b$ output
qubits is a superoperator $Q: \linear(\Hspace^{\otimes
a})\to\linear(\Hspace^{\otimes b})$. There are certain constraints
that $Q$ must satisfy in order to represent a physically realizable
quantum operation. We need not be concerned about those constraints
as our simulation method does not depend on them.
In existing applications one typically has $a\geq b$ and often $a=b$,
though a density operator can also be regarded as a gate with $a=0$.
The ordering of inputs and
outputs is in general significant. If $Q$ is a {\em traced out}
operator, then $b=0$, and $Q(|x\rangle\langle y|)=\langle x |
y\rangle$, for all $x, y\in\{0, 1\}^a$.
We denote by $Q[A]$ the application of $Q$ to an ordered set $A$
of $a$ qubits.

The information in a quantum state is retrieved through the application
of measurements. A POVM (Positive Operator-Valued Measure) $\mathcal{M}$
on $n$ qubits is a set
$\mathcal{M}=\{M_1, M_2, \cdots, M_k\}$, where each $M_i$
is called a POVM element, and is a positive semi-definite operator
in $\linear(\Hspace^{\otimes n})$ such that $\sum_{i=1}^k M_i=I$.
The single-qubit measurement in the computational basis is
$\{|0\rangle\langle 0|, |1\rangle\langle 1|\}$.

We assume that the maximum number of qubits on which a quantum gate can act
is bounded by a constant (often two or three).
A quantum circuit of size $T$ with $n$ input-qubits and
$m$ output-qubits consists of the following:
\begin{enumerate}[(1)]
\item A sequence of $n$ input-wires, each of
which represents one input-qubit, i.e., a qubit
which is not the output qubit of any gate. 
\item A sequence of $T$ quantum gates
$g_1$, $g_2$, $\ldots$, $g_T$, each of which is applied to some subset of the wires. 
\item A sequence of $m$ output-wires, each of which represents
an output-qubit, i.e., a qubit which is not the input qubit of any gate.
\end{enumerate}
Note that by the above definition,
a quantum circuit $C$ defines a function $C: \linear(\Hspace^{\otimes n})
\rightarrow \linear(\Hspace^{\otimes m})$.
In most applications, a circuit $C$ is applied to 
an input state $\rho_x\defeq \otimes_{i=1}^n |x_i\rangle\langle x_i|$,
for some binary string $x=x_1\cdots x_n\in\{0, 1\}^n$,
and at the end of the computation, measurements in the computational basis 
are applied to a subset of the qubits. We shall restrict our discussions
to such case, though our results can be extended to more general cases.

The graph of a quantum circuit $C$, denoted by $G_C$, is obtained from $C$ as
follows. Regard each gate as a vertex, and for each input/output wire add a new
vertex to the open edge of the wire.\footnote{These vertices are going to 
represent input states, as well as measurements and trace-out operators at
the end of the computation.}
Each wire segment can now be represented by an edge in the graph.
\end{trivlist}
\section{Tensors and tensor networks}
Tensors, commonly used in physics, are multi-dimensional
matrices that generalize more traditional tools from linear algebra,
such as matrix products. Here we focus on features of tensors that are
relevant to our work.

\begin{definition}
A rank-$k$ tensor in an $m$-dimension space
$g=[g_{i_1, i_2, \ldots, i_k}]_{i_1, i_2, \ldots, i_k}$
is an $m^k$-dimensional array of complex numbers $g_{i_1, i_2, \ldots, i_k}$,
indexed by $k$ indices, $i_1$, $i_2$, $\ldots$, $i_k$, each of which takes
$m$ values.  When the indices are clear we omit them outside the bracket.
\end{definition}
 
For example, a rank-$0$ tensor is simply a complex number, and a rank-$1$ tensor
is a dimension-$m$ complex vector. We focus on dimension-$4$ tensors,
and set the range of each index to be $\Pi\defeq\{ |b_1\rangle\langle b_2| :
 b_1, b_2\in \{0, 1\}\}$.
We fix the following tensor representation of
a density operator and a superoperator.

\begin{definition}
Let $\rho$ be a density operator
on $a$ qubits. The tensor of $\rho$
is  $[ \rho_{\sigma_1, \sigma_2, \ldots, \sigma_a}]_{\sigma_1, \sigma_2, \ldots, \sigma_a\in \Pi}$,
where \[\rho_{\sigma_1, \ldots, \sigma_a}\defeq tr(\rho\cdot (\otimes_{i=1}^a \sigma_i)^\dagger).\]

Let $Q$ be a superoperator acting on $a$ input qubits and $b$ output qubits.
The tensor of $Q$ is \[Q_{\sigma_1, \sigma_2, \cdots, \sigma_a,
     \tau_1, \tau_2, \cdots, \tau_b}]_{\sigma_1, \ldots, \sigma_a, \tau_1,
     \ldots, \tau_b\in\Pi},\] where
     \[Q_{\sigma_1, \sigma_2, \cdots, \sigma_a, \tau_1, \tau_2, \cdots, \tau_b}
     \defeq tr(Q(\otimes_{i=1}^a \sigma_i)\cdot (\otimes_{j=1}^b \tau_j)^\dagger).
     \]
\end{definition}

We shall use the same notation for a density operator (or a superoperator)
and its tensor. We now define the central object of the paper.
\begin{definition}
A {\em tensor network} is a
collection tensors, each index of which may be used by either one or two tensors.
\end{definition}

\begin{figure}[tbph]
\centering
\includegraphics[width=6in]{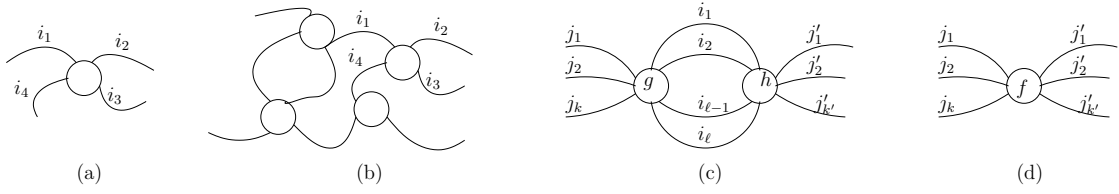}
\caption{A rank-4 tensor is illustrated in (a),
and a tensor network with four tensors is shown in (b).
Contraction of two tensors is illustrated in (c) and (d).}
\label{fig:tensor}
\end{figure}

A rank-$k$ tensor $g$ can be graphically represented as a vertex labeled with
$g$, and connected to $k$ open wires, each of which is labeled with a distinct
index. We may represent a tensor network
by starting with such graphical representations of its tensors,
and then connecting wires corresponding to the same index.
Note that now each wire corresponds to a distinct index.
Also, an index that appears in one tensor corresponds to an open wire,
and an index that appears in two tensors corresponds to an edge
connecting two vertices.
Parts (a) and (b) in Figure~\ref{fig:tensor} give an example
of the graphical representation of a tensor and a tensor network.
In the tensor $g_Q$, we call the $\sigma_i$ wires,
$1\le i\le a$, input wires, and the $\tau_j$ wires, $1\le j\le b$, the output wires.

Suppose in a tensor network, 
there are $\ell$ parallel edges
$i_1$, $i_2$, $\ldots$, $i_\ell$ between 
two vertices $g=[g_{i_1, \ldots,
i_\ell, j_1, \ldots, j_k}]$ and $h=[h_{i_1, \ldots, i_\ell, j'_1, \ldots,
j'_{k'}}]$.  We may contract those edges by first removing them,
then merging $v_g$ and $v_h$ into a new vertex $v_{f}$, whose
tensor is $f=[ f_{j_1, \ldots, j_k, j'_1, \ldots, j'_{k'}} ]$, and
\begin{equation}
\label{eqn:tensorContraction}
  f_{j_1, \ldots, j_k, j'_1, \ldots, j'_{k'}}
 \ \defeq\
\sum_{i_1, i_2, \ldots, i_\ell} g_{i_1, \ldots, i_\ell, j_1,\ldots, j_k}\ \cdot\
 h_{i_1, \ldots, i_\ell, j'_1, \ldots, j'_{k'}}.
\end{equation}

\begin{figure}[tbph]
\centering
\includegraphics[width=3in]{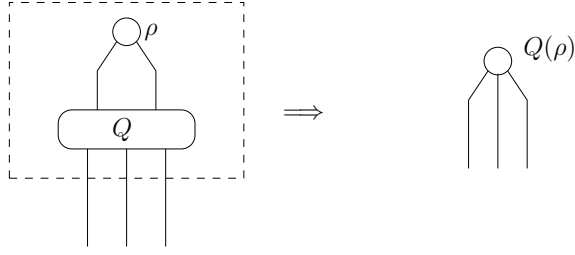}
\caption{Contracting the wires connecting the tensors
for a density operator $\rho$ and a gate $Q$ results
in the tensor for $Q(\rho)$.}
\label{fig:tensor-gate}
\end{figure}

Parts (c) and (d) in Figure~\ref{fig:tensor} illustrate the above contraction.
Note that a tensor network with $k$ open wires can be contracted to a single tensor
of rank $k$, and the result does not depend on the order of contractions.
The following example is instructive.
\begin{example} Let $\rho$ be an $a$-qubit density operator and
$Q$ be a superoperator with $a$ input qubits and $b$ output qubits.
Consider the tensor network that connects all wires of the tensor $\rho$ to
the input wires of the tensor $Q$. Then contracting this tensor network
gives the tensor of the density operator $Q(\rho)$.
Figure~\ref{fig:tensor-gate} illustrates this example.
\end{example}

\begin{figure}[tbph]
\centering
\includegraphics[width=6in]{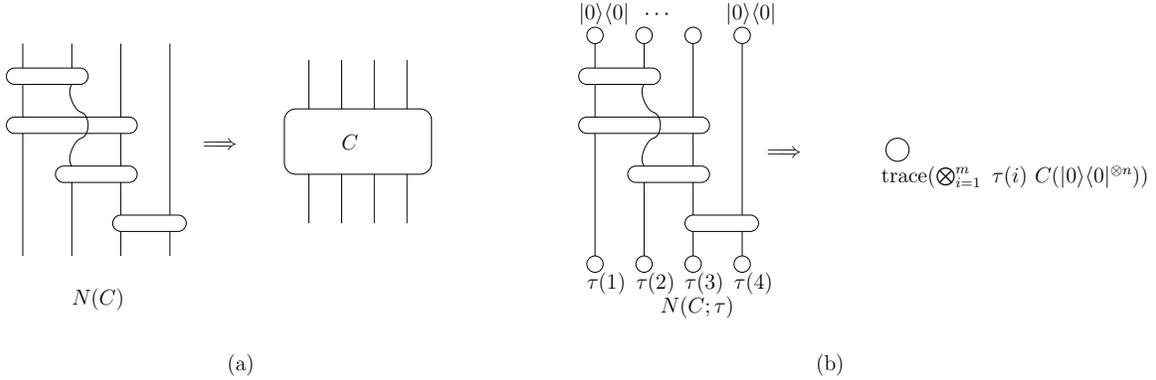}
\caption{In (a), a circuit $C$ can be naturally regarded
as a tensor network $N(C)$. Contracting $N(C)$ gives
the tensor for the operator that $C$ realizes.
Part (b) illustrates the tensor network $N(C; \tau)$,
contracting which gives the rank-$0$ tensor
whose value is precisely the probability that the measurement scenario 
$\tau$ is realized on $C(|0\rangle\langle0|^{\otimes n})$.}
\label{fig:circuit}
\end{figure}

A quantum circuit $C$ can be naturally regarded
as a tensor network $N(C)$:
each gate is regarded as the corresponding tensor.
The qubit lines are wires connecting the tensors,
or open wires that correspond to the input and output qubits.
Figure~\ref{fig:circuit} illustrates the concept.

Let $C$ be a quantum circuit with $n$ input qubits and $m$ output qubits.
Suppose that $C$ is applied to the initial state $\rho_x$, for some $x\in\{0, 1\}^n$,
and we are interested in knowing the probability of observing
some particular outcome when some single-qubit measurements
are applied to a subset of the qubits. The setting
can be described by a measurement scenario defined as follows.
\begin{definition}
\label{def:measurementScenario}
Let $m\ge1$ be an integer. A {\em measurement scenario}
on $m$ qubits is a function
$\tau:[m]\to \linear(\mathbb{C}^2)$, \label{sec:measurementScenario}
such that $\tau(i)$ is a single-qubit POVM measurement element.
\end{definition}
Note that if a qubit $i$ is not to be measured, we can set
$\tau(i)=I$.

To compute the probability that $\tau$ is realized on $C(\rho_x)$,
we build a tensor network $N(C; x, \tau)$ from $N(C)$ by
attaching to each input open wire $i$  the tensor
for $|x_i\rangle\langle x_i|$, and attaching
to each open wire for the output qubit $i$
the tensor for $\tau(i)$.
When $x=0^n$, we abbreviate $N(C; x, \tau)$ as $N(C; \tau)$.
Figure~\ref{fig:circuit} illustrates the concept of $N(C)$ and $N(C; \tau)$.

\begin{proposition}\label{res:calcProb}
Let $C$ be a quantum circuit, $x$ be a binary string, and $\tau$ be a measurement scenario.
Contracting the tensor network $N(C; x, \tau)$ to a single vertex gives the
rank-$0$ tensor which is the probability that $\tau$ is realized on $C(\rho_x)$.
\end{proposition}

\begin{proof} Let $\rho^t\defeq g_t g_{t-1}\cdots g_1(\rho_x)$, $1\le t\le T$,
and $\rho^0=\rho_x$. By the definitions of tensors for density operators
and superoperators and tensor contraction, contracting wires connecting
the tensor of a superoperator $Q$ and the tensors for a density operator $\rho$
gives the tensor of $Q(\rho)$. Thus sequentially contracting input wires of
$g_1$, $\cdots$, $g_t$ gives the tensor for $\rho^t$, and contracting
the remaining wires gives the tensor for $\tau(\rho^T)$, which is the
probability of realizing $\tau$ on $\rho^T=C(\rho_x)$. 
\end{proof}

We remark that $N(C; x, \tau)$ is not the only tensor network
for which the above Proposition holds.

Although the ordering of the edges in the contraction process does not
affect the final tensor, it may significantly affect space and time 
requirements.

\begin{proposition}\label{res:runtime}
Given a tensor network $N$ of a
size $T$ quantum circuit,
and a contraction process specified by an ordering of
wires in $N$, let $d$ be the maximum rank of
all the tensors that appear in the process.
Then the contraction takes $O(T\exp[O(d)])$ time.
\end{proposition}

\begin{proof}
Note that the size of $N$ is $\Theta(T)$.  The algorithm stores the
tensors of each vertex.  When contracting an edge, it computes the new tensor
according to Equation \ref{eqn:tensorContraction}, and updates the tensor
accordingly. This takes $\exp[O(d)]$ time. Hence the total runtime is
$O(T \exp[O(d)])$.
\end{proof}

In the next Section we will investigate near-optimal orderings for
simulation and ways to find them. While traditional simulation of
quantum circuits proceeds in the same order in which the gates are
applied, it appears that an optimal ordering may not have any 
physical meaning. Therefore, we formalize this optimization 
using abstract graph contractions. 

\section{Contraction complexity and treewidth}
Let $G$ be a graph with vertex set $V(G)$ and edge set $E(G)$.
Recall that the contraction process discussed in the previous
Section removes parallel edges in one step because contracting
one edge at a time can create multiple loops. However, for 
future convenience we prefer the latter simulation and therefore
allow loops to remain not contracted, counting toward the degree
of a vertex. Note that if a ``parallel'' contraction
contracts $\ell$ edges between two vertices $u$ and $v$ of 
degrees $\ell+k$ and $\ell+k'$, respectively,
the corresponding ``one-edge-at-a-time'' contraction
would create vertices of degrees $k+k'+\ell-1$, $k+k'+\ell-2$,
$\cdots$, $k+k'$, each of which is $\le \degree(u)+\degree(v)$.
Thus the one-edge-at-a-time contraction process 
can emulate the parallel contraction, while increasing
the maximum vertex degree observed by no more than two-fold.
We make the definition of this new contraction process precise
below.

\begin{definition}
The {\em contraction of an edge} $e$ removes $e$ and
replaces its end vertices (or vertex) with a single
vertex. A {\em contraction ordering} $\pi$ is an
ordering of all the edges of $G$, $\pi(1)$, $\pi(2)$, $\ldots$,
$\pi(|E(G)|)$. 
The complexity of $\pi$
is the maximum degree of a merged vertex during the contraction process. 
The {\em contraction complexity} of $G$, denoted by $\cc(G)$, 
is the minimum complexity of a contraction ordering.
\end{definition}
Since only the degrees of the merged vertices are considered
in defining the contraction complexity, $\cc(G)$ could be strictly
larger than $\maxDegree(G)$. For example, if $G$ is a path, $\cc(G)=1$ and 
$\maxDegree(G)=2$.

Note that sequentially contracting all $\pi(i)$, $1\le i\le
|E(G)|$, reduces $G$ to a single vertex (or an empty graph
of several vertices). 
Also, for any graph $G$, $\cc(G)\le |E(G)|-1$, since any merged vertex
would be incident to no more than $|E(G)|-1$ number of edges.
Furthermore, $\cc(G)\ge \maxDegree(G)-1$, since when an edge incident to
a vertex of degree $\maxDegree(G)$ is removed, the resulting merged vertex
is incident to at least $\maxDegree(G)-1$ edges.

The nature of $\cc(G)$ becomes clearer once we consider
the {\em line graph} of $G$, denoted by $G^*$.
That is, the vertex set of $G^*$ is
$V(G^*) \defeq E(G)$, and the edge set is
\[E(G^*)\defeq
\{ \{ e_1, e_2\}\subseteq E(G) : \textrm{$e_1\ne e_2$, $\exists v\in V(G)$ such that
$e_1$ and $e_2$ are both incident to $v$}\}.\]

\begin{proposition}\label{res:cctw}
For any graph $G=(V, E)$, $\cc(G)=\tw(G^*)$.
Furthermore, given a tree decomposition of $G^*$ of width $d$,
there is a deterministic algorithm that outputs
a contraction ordering $\pi$ with $\cc(\pi)\le d$
in polynomial time.
\end{proposition}

Computing the treewidth of an arbitrary graph is NP-hard \cite{ArnborgCP87},
but we do not know if this remains true for the special class
of graphs $G^*$. Nevertheless, this is not critical in our work since
the constant-factor approximation due to Robertson and Seymour~\cite{RSX} 
suffices for us to prove our key results.

\begin{theorem}[Robertson and Seymour~\cite{RSX}]\label{res:rs}
There is a deterministic algorithm
that given a graph $G$ outputs
a tree decomposition of $G$ of width $O(\tw(G))$ in
time $|V(G)|^{O(1)} \exp[O(\tw(G))]$.
\end{theorem}

\begin{proofof}{Proposition~\ref{res:cctw}}
There is a one-to-one correspondence of the contraction of an edge
in $G$ and the elimination of a vertex in $G^*$, and the degree
of the merged vertex resulting from contracting an edge $e$ in $G$ is
the same as the degree of $e$ being eliminated in $G^*$.
Thus $\cc(G)=\tw(G^*)$.

To prove the second part of the statement, 
denote the tree decomposition
by $\mathcal{T}$. Repeat the following until
the tree decomposition becomes an empty graph.
Choose a leaf $\ell$ in $\mathcal{T}$.
If $\ell$ is the single vertex of $\mathcal{T}$, output
vertices (of $G^*$) in $B_\ell$ in any order.
Otherwise, let $\ell'$ be its parent.
If $B_\ell\subseteq B_{\ell'}$, remove
$\ell$ and repeat this process.
Otherwise, let $e\in B_{\ell}-B_{\ell'}$.
Output $e$, remove it from the tree decomposition
and continue the process, until all vertices
of the tree decomposition are removed.
The number of steps in this process is polynomial in the size of the
tree decomposition.

Note that each output $e$ appears in only one bag
in the tree decomposition. Therefore, all (current)
neighbors of $e$ must appear in the same bag. Hence
its induced width is at most $d$. By the one-to-one
correspondence of the vertex elimination in $G^*$
and the contraction process in $G$, $\cc(\pi)\le d$.
\end{proofof}

Before we complete the description of our simulation algorithm,
we relate the treewidth of $G$ to that of $G^*$.
This is useful for reasoning about quantum circuits $C$ when
the graph $G_C$ is easier to analyze than its line graph $G^*_C$.
In such cases one hopes to bound the runtime of the simulation algorithm
in terms of parameters of $G$ rather than $G^*$.
Fortunately, since $G_C$ is of bounded degree,
the treewidths of $G_C$ and $G^*_C$ are asymptotically the same.

\begin{lemma}\label{res:sameTreeWidth}
For any graph $G$ of maximum degree $\maxDegree(G)$,
\[ (\tw(G)-1)/2\le \tw(G^*)\le \maxDegree(G)(\tw(G)+1)-1.\]
\end{lemma}
\begin{proof}
 From a tree decomposition $\mathcal{T}$ of $G$ of width $d$ we
obtain a tree decomposition $\mathcal{T}^*$ of $G^*$ of width
$(d+1)\cdot \maxDegree(G)-1$ by replacing each vertex $v\in V(G)$ with all edges
$e$ incident to $v$. This guarantees that every edge of
$G^*$ is in some bag, i.e. (T1) is true. Item (T2) is true
since if $e_1$ and $e_2$ are both incident to a vertex $u$ in $G$,
then any bag in $\mathcal{T}$ containing $u$ contains both $e_1$
and $e_2$ in $\mathcal{T}^*$. 
To verify Item (T3), suppose that $e$ connects $u$ and $v$ in $V(G)$.
Take two bags $a$ and $b$ that both contain $e$.
Then in $\mathcal{T}$,  both bags $a$ and $b$ must have either
$u$ or $v$. If they contain the
same vertex, then $a$ and $b$ are connected, by (T3). Otherwise,
there must be a bag $c$ that contains both $u$ and $v$, by (T2).
So $a$ and $b$ are connected through $c$.
Therefore we have proved that $\tw(G^*)\le \maxDegree(G)(\tw(G)+1)-1$.

Now to prove $\tw(G)\le 2\tw(G^*)+1$, we start with
a tree decomposition $\mathcal{T}^*$ of $G^*$  of width
$d$, and replace every $e$ by its two end vertices in $V(G)$.
The verification of (T1) through (T3) can be accomplished in a similar way.
\end{proof}

Note that the above bounds are asymptotically tight, since
for an $m$-ary tree (of which each non-root internal vertex has degree $m+1$),
the treewidth is $1$ and the contraction complexity is $m$. 
We summarize the above finding in the following theorem.

\begin{theorem}\label{res:equivalence}
Let $d\ge1$ be an integer.
For any family of graphs $G_n$, $n\in\mathbb{N}$,
such that $\maxDegree(G_n)\le d$, for all $n$,
then
\[  (\tw(G_n)-1)/2\le \cc(G_n)=\tw(G^*_n) \le d(\tw(G_n)+1)-1,\quad\forall
n\in\mathbb{N}.\]
\end{theorem}

We are now ready to put everything together
to prove the following restatement of Theorem~\ref{res:main}.

\begin{theorem}\label{res:simuCCTW}
Let $C$ be a quantum circuit of size $T$ and
with $n$ input and $m$ output qubits, $x\in\{0, 1\}^n$ be an input,
and $\tau:[m]\to\linear(\mathbb{C}^2)$ be a
measurement scenario. Denote by $G_C$ the underlying
circuit graph of $C$. Then the probability that 
$\tau$ is realized on $C(\rho_x)$ can be computed
deterministically in time $T^{O(1)}\exp[O(\cc(G_C))]
=T^{(1)}\exp[O(\tw(G_C))]$.
\end{theorem} 

\begin{proof}
The following algorithm computes the
desired probability.

\begin{enumerate}[(1)]
\item Construct $N=N(C; x, \tau)$.
\item \label{step:rs} Apply the Robertson-Seymour algorithm to compute
a tree decomposition $\mathcal{T}$ of ${N}^*$
of width $w=O(\tw({N}^*))$
(Theorem \ref{res:rs}).
\item Find a contraction ordering $\pi$ from $\mathcal{T}$
(Proposition \ref{res:cctw}) of width $w$.
\item \label{step:calProp} Contract ${N}$ using $\pi$, and
output the desired probability from the final (rank-$0$) tensor
(Proposition \ref{res:calcProb}).
\end{enumerate}

The runtime bottlenecks are Steps (\ref{step:rs}) and (\ref{step:calProp}),
both taking time $T^{O(1)}\exp(O[\tw({N}^*)])$, which by Theorem \ref{res:equivalence}
is $T^{O(1)}\exp[O(\cc(G_C))] = T^{O(1)}\exp[O(\tw(G_C))]$. In
fact, Steps 2 and 4 can be combined, but we separate them for the
sake of clarity.
\end{proof}

\section{Treewidth and quantum circuits}
In this section we prove the implications of Theorem~\ref{res:main}
stated in the Introduction.
A number of tight bounds for the treewidth of specific families of graphs
have been published, including those for planar and series-parallel graphs.
However, similar results for graphs derived from quantum circuits are lacking.
To this end, we strengthen Corollary \ref{res:near} as follows.

\begin{proposition}\label{res:logdepth}
Let $C$ be a quantum circuit in which each gate
has an equal number of input and output qubits,
and whose qubits are index by $[n]$, for an integer
$n\ge1$. Suppose that the size of $C$ is $T$, and $r$ is the minimum
integer so that for any $i$, $1\le i\le n-1$, no
more than $r$ gates act on some qubits $j$ and $j'$ with $j\le i
<j'$. Then $C$ can be simulated deterministically in time $T^{O(1)}\exp[O(r)]$.
\end{proposition}

Corollary~\ref{res:near} follows since $r=O(qD)$ under its assumption.

\begin{proofof}{Proposition~\ref{res:logdepth}}
Assume without loss of generality that $\tw(G_C)\ge2$.
Let $G$ be the graph obtained from $G_C$ by removing
degree $1$ vertices and contracting edges incident to degree $2$ vertices.
Then $\tw(G)=\tw(G_C)$, by Proposition \ref{res:insert} and the observation stated
before it. Then each vertex in $G$ corresponds to a multi-qubit gate in $C$.

We now construct a tree decomposition $\mathcal{T}$ for $G$
that forms a path of $n-1$ vertices $B_1 \!-\! B_2\! - \cdots -\! B_{n-1}$. 
The bag $B_i$ of the $i^{th}$ vertex ($1\le i\le
n-1$) consists of multi-qubit gates (vertices) that act on some qubits $j$ and
$j'$ with $j\le i<j'$. Hence $|B_i|\le r$ by the assumption.
If $u$ acts on qubits $i_1, i_2, \cdots, i_k$, $i_1<i_2<\cdots<i_k$,
then $u\in B_i$, for all $i$, $i_1\le i\le i_k$. Thus (T1) and (T3)
are true.
If a wire segment corresponding to the qubit $i$ connects two gates
$u$ and $v$, the bag $B_i$ contains both $u$ and $v$. Thus (T2) is true.
Therefore $\mathcal{T}$is a tree decomposition for $G$ with width $r-1$.
Hence $\tw(G_C)=\tw(G)=O(r)$, which by Theorem \ref{res:main} implies that
$C$ can be simulated in $T^{O(1)}\exp[O(r)]$ time.
\end{proofof}

We now turn to quantum circuits of bounded depth.
To prove Theorem \ref{res:depth4} we will make use of the following
observation that relates expander graphs to contraction complexity.
Let $d$ be a constant and $\{G_n\}_{n\in\mathbb{N}}$ be a family of
$d$-regular graphs, and $\epsilon>0$ be a universal constant.
Recall that 
 $\{G_n\}$ is called a family of expander graphs with expansion
 parameter $\epsilon$ if, for any subset $S\subseteq V(G_n)$ with
 $|S|\le |V(G_n)|/2$, there are no less than $\epsilon |S|$
 edges connecting vertices in $S$ with vertices in $V(G)-S$.
\begin{lemma}\label{res:expander}
For an expander graph $G_n$ with the expansion parameter $\epsilon$,
$\cc(G_n)\ge \epsilon|V(G_n)|/4$.
\end{lemma}
\begin{proof} Fix a contraction ordering of $G_n$.
Let $v$ be the first merged vertex so that $k_v$, the number
of vertices in $V(G_n)$ that were eventually merged to $v$, is at least
$|V(G_n)|/4$. Then $k_v\le |V(G_n)|/2$, and
$v$ must have degree $\epsilon |V(G_n)|/4$.
\end{proof}

The following graph is shown to be an expander by
Lubotzky, Phillips and Sarnak~\cite{LubotzkyPS88}.
Let $p>2$ be a prime, and $G_n$ be the graph
with $V(G_p)\defeq \mathbb{Z}_p\cup \{\infty\}$,
and every vertex $x$ is connected to $x+1$, $x-1$ and $x^{-1}$
($\infty\pm 1$ are defined to be $\infty$).
Note that $G_p$ is a $3$-regular graph.

\begin{proofof}{Theorem \ref{res:depth4}}
By Lemma \ref{res:expander}, $\cc(G_p)=\Omega(p)$.
Since $G_p$ is a $3$ regular graph,
$\tw(G_p)=\Theta(\cc(G_p))=\Omega(p)$, by Theorem
\ref{res:equivalence}. Let $G'_p$ be the graph
obtained from $G_p$ by
removing the vertex $\infty$ and the edge $\{0, p-1\}$.
This would only decrease $\tw(G_p)$ by at most constant.
Hence $\tw(G'_p)=\Omega(p)$.
Therefore to prove the theorem,
it suffices to construct a quantum circuit $C$ on $p$
qubits so that $G'_p$ is a minor of $G_C^*$.

Each qubit of $C$ corresponds to a distinct vertex in $V(G'_p)$.
Observe that edges in $E(G'_p)$ can be partitioned into three
vertex-disjoint subsets: (1) $\{x, x^{-1}\}$; (2) $\{x, x+1\}$ for
even $x$, $0\le x\le p-3$; (3) the remaining edges. Each subset
gives a layer of two-qubit gates in $C$. In $G^*_C$, contracting all
the vertices that correspond to the same qubit gives a graph of
which $G'_p$ is a minor. Hence
$\tw(C)=\Theta(\tw(G_C^*))=\Omega(p)$.
\end{proofof}

\begin{proofof}{Theorem \ref{res:depth3}}
By Theorem \ref{res:equivalence},
it suffices to prove that $\cc(G_C)= O(1)$ for any depth-2 circuit.
Observe that for any such circuit, after contracting
the input and output vertices (those are of degree $1$, hence
contracting them will not increase the contraction complexity),
every vertex in $G_C$ has degree either $1$ or $2$.
Hence the edges can be decomposed into disjoint paths
and cycles, which can be contracted without increasing the degree.
Hence $\cc(G_C)\le 2$.
\end{proofof}

\section{Simulating one-way quantum computation}
\label{sec:oneway}
This section revisits the notions of {\em graph states} and 
{\em one-way quantum computation}. We first 
simulate one-way computation with an algorithm
whose complexity grows exponentially with the contraction complexity
of the underlying graph. We then reduce general one-way computation 
to the special case where the vertex degree is bounded by a constant.
Since for such graphs the contraction complexity is the same 
as the treewidth (up to a constant), this reduction facilitates
a more efficient simulation algorithm, as stated in Theorem~\ref{res:oneway}.

Let $G=(V, E)$ be a simple undirected graph 
with $|V|=n$. For a subset $V'\subseteq V$, denote by $e(V')$ the number
of edges in the subgraph induced by $V'$. 
We associate a qubit with each vertex $v\in V$, and refer to it by 
qubit $v$. For a subset $V'\subseteq V$, we identify
the notation $|V'\rangle$ with the computational basis
$|x\rangle$, for $x\in\{0, 1\}^n$ being the characteristic
vector of $V'$ (i.e., the $i^{th}$ bit of $x$ is $1$ if and only
if the $i^{th}$ vertex under some fixed ordering is in $V'$).
The graph state $|G\rangle$ is the following
$n$-qubit quantum state~\cite{BR}
\[ |G\rangle\defeq \frac{1}{\sqrt{2^n}}\
 \sum_{V'\subseteq V} (-1)^{e(V')}|V'\rangle.\]
Note that $|G\rangle$ can be created from $|0^n\rangle$
by first applying Hadamard gates to all qubits, followed
by the Controlled-Phase gate $\Lambda(\sigma^z) = \sum_{b_1, b_2\in\{0, 1\}}
(-1)^{b_1\cdot b_2} |b_1, b_2\rangle\langle b_1, b_2|$
on each pair of qubits $u$ and $v$ with $\{u, v\}\in E$.
Since all the $\Lambda(\sigma^z)$ operators commute, the order of applying
them does not affect the result.

A basic building block of our simulation algorithm is the following.
\begin{lemma}\label{res:singleQubitMeasure}
Let $G=(V, E)$ be a graph with $n$ vertices, and $\tau$ be a measuring
scenario (defined in Definition~\pageref{sec:measurementScenario})
 on $n$ qubits. 
Then the probability $p$ that $\tau$ is realized on $|G\rangle$
can be computed deterministically in time $O(|V|^{O(1)}\exp[O(\cc(G))])$.
\end{lemma}
\begin{proof}
Fix a circuit $C_G$ that creates $|G\rangle$ from $|0\rangle\langle0|^{\otimes n}$.
Let $\{u, v\}\in E$, and 
$g=g_{u^+, u^-, v^+, v^-}$ be a tensor in $N(C_G; \tau)$
corresponding to $\Lambda(\sigma^z)[u, v]$.
The wires representing the qubit $u$ (or $v$) before and after
the gate are labeled $u^+$ (or $v^+$) and $u^-$ (or $v^-$), respectively.
We replace $g$ by two tensors $g^u=g^u_{u^+, u^-, t^+, t^-}$
and $g^v=g^v_{v^+, v^-, t^+, t^-}$, which share two labels $t^+$ and $t^-$
and are defined as follows. For a wire segment with a label $a$,
denote by $\linear_a$ the $4$-dimensional space of linear operators associated with
this wire segment. Set
$g^u$ to be the identity superoperator
that maps $\linear_{u^+}\otimes \linear_{t^-} \to \linear_{t^+}\otimes\linear_{u^-}$,
and $g^v$ to be the tensor for a $\Lambda(\sigma^z)$ that maps
$\linear_{t^+}\otimes\linear_{v^+} \to \linear_{t^-}\otimes\linear_{v^-}$.
By their definitions, contracting $g^u$ and $g^v$ gives precisely $g$.
We call the inserted wires labeled with $t^+$ and $t^-$ {\em transition wires}.
See Figure~\ref{fig:replaceSigmaZ} for an illustration.

\begin{figure}[tbph]
\centering
\includegraphics[width=3in]{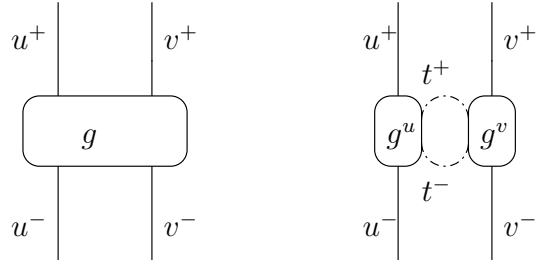}
\caption{Replacing a tensor $g$ corresponding to
$\sigma_z[u, v]$ by two tensors $g^u$ and $g^v$.}
\label{fig:replaceSigmaZ}
\end{figure}

Denote by $N'(C_G; \tau)$ the tensor network obtained from $N(C_G; \tau)$ by applying
the above replacement procedure for each edge in $E$. Let $G'$ 
be the underlying graph of $N'(C_G; \tau)$. Note that
$G'$ has the maximum degree $4$ and the number of vertices is $O(|E|)$.
See Figure~\ref{fig:one_measurement} for an illustration.
Thus $p$ can be computed
by contracting $N'(C_G; \tau)$ in time $O(|V|^{O(1)}\exp[O(\cc(G'))])$,
according to Theorem~\ref{res:simuCCTW}. 

\begin{figure}[tbph]
\centering
\includegraphics[width=6in]{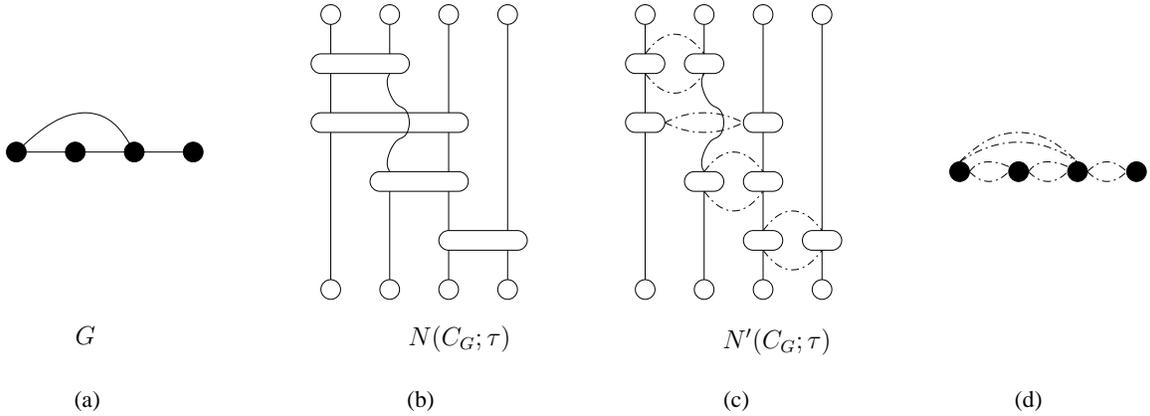}
\caption{For a graph $G$ in (a), the tensor network $N(C_G; \tau)$
is shown in (b). Input vertices are at the top, and output vertices
are at the bottom. Each box
is a tensor corresponding to a $\Lambda(\sigma_z)$ applied to qubits adjacent
in $G$. In (c), each $\Lambda(\sigma_z)$ tensor is replaced by two tensors
and two wires connecting them, as described in Figure~\ref{fig:replaceSigmaZ}.
Contracting all solid lines in (c) produces the graph in (d),
which is precisely $G$ with each edge doubled.}
\label{fig:one_measurement}
\end{figure}

We now prove that $\cc(G')=O(\cc(G))$.
This can be seen by contracting all wire segments corresponding to the same
qubit in $G'$, while leaving the transition wires untouched.
Since contracting the edge incident to an input or output vertex results
in a new vertex of degree $3$, and contracting the rest of the wires
for a qubit $v$ results in a new vertex of degree $2\degree(v)$,
the maximum degree of a merged vertex in this process is $\max \{3, 2 \maxDegree(G)\}$.
The one-to-one correspondence between the resulting vertex set 
and $V$ induces naturally a one-to-one correspondence between
the pairs of transition wires and $E$. Thus a contraction ordering of $G$
gives a contraction ordering of $G'$ (of this stage)
with at most twice of the contraction complexity.
Therefore
\[ \cc(G')\le \max \{3, 2\maxDegree(G), 2\cc(G)\} = O(\cc(G)+1).\]
Thus $p$ can be computed deterministically in time 
 $O(|V|^{O(1)}\exp[O(\cc(G'))])= O(|V|^{O(1)}\exp[O(\cc(G))])$.
\end{proof}

A {\em one-way} computation on a quantum state $|\phi\rangle$
consists of a sequence of adaptive single-qubit measurements
and single-qubit unitary operations applied to $|\phi\rangle$.
The description of each measurement or unitary operation, including the index of the qubit that it 
acts on, can be computed by a deterministic and efficient (polynomial time) algorithm
from previous operations and their measurement outcomes.
In our discussion we treat this computation time as a constant.
We call a one-way quantum computation {\em oblivious} if before the last
measurement (which produces the outcome of the computation), different
computational paths involve the same number of measurements,
take place with the same probability, and result in
an identical state.
Note that the one-way computation of Raussendorf and Briegel~\cite{RB}
is oblivious.

We point out that allowing single-qubit unitary operations in the definition
is for the convenience of discussion only, since each single-qubit
unitary can be combined with a future measurement on the same qubit
(should there be one). 
To see this fact, let us call two quantum states {\em LU-equivalent} (where LU stands for
Local Unitary), if there exists a set of single-qubit unitary operations
applying which maps one state to the other.
A one-way computation with unitary operators always has an almost identical
one-way computation without unitary operations:
the measurements are in one-to-one correspondence with identical
outcome distributions, and the states after corresponding measurements
are LU-equivalent. Therefore, when we are only interested in the
distribution of the measurement outcomes, we may assume without loss of generality
that a one-way computation does not involve any unitary operation. 

We now derive a simulation algorithm whose complexity depends
on the contraction complexity.
\begin{lemma}\label{res:onewaydegree}
A one-way quantum computation on a graph $G=(V, E)$ can be simulated
by a randomized algorithm in time $O(|V|^{O(1)}\exp[O(\cc(G)])$.
If the one-way computation is oblivious, the simulation can be made deterministic.
\end{lemma}

\begin{proof}
Let $T$ be the number of measurements during the one-way computation.
Assume without loss of generality that no single-qubit 
unitary operation is applied.
The simulation consists of $T$ steps, one for each single-qubit measurement.
It maintains a data structure $r=(\tau, p)$,
where $\tau$ is a measurement scenario, and $p$ is the probability
that $\tau$ is realized on $|G\rangle$. 
Denote by $r_t=(\tau_t, p_t)$ the value
of $r$ when $t$ measurements have been simulated.
Initially $\tau_0(i)=I$ for all $i$, $1\le i\le n$, and $p_0=1$.

Suppose we have simulated
the first $t-1$ measurements, $1\le t\le T-1$.
\begin{enumerate}[(1)]
\item Based on the one-way algorithm,
compute from $\tau_{t-1}$ the description of the $t^{th}$ measurement
$P_t=\{P_t^0, P_t^1\}$ and the qubit $a_t$
that it acts on.
Denote by $\tau_t^0$ the measurement scenario identical to $\tau_{t-1}$,
except that $\tau_t^0(a_t)= P_t^0$.
\item Compute $p_t^0$, the probability of realizing $\tau_t^0$.
By Lemma~\ref{res:singleQubitMeasure}, this takes $O(|V|^{O(1)}\exp[O(\cc(G))])$
time.
\item Flip a coin that produces $0$ with probability $p_t^0/p_{t-1}$,
resulting in an outcome $b_t\in\{0, 1\}$.
Set $\tau_t$ to be identical to $\tau_{t-1}$,
except that $\tau(a_t)=p_t^{b_t}$. Set $p_t=(1-b_t)p_t^0 + b_t(p_{t-1}-p_t^0)$. 
Continue the simulation until $t=T$.
\end{enumerate}

By construction, the output distribution is identical to that 
of the one-way computation. The complexity of the algorithm is
$O(|V|^{O(1)}\exp[\cc(G)])$. 

If the one-way computation is oblivious, 
there is no need to adaptively simulate the first $T-1$ measurements,
as all of them lead to the same state with the same probability $p_{T-1}$.
Let $\tau_{T-1}$ ($\tau_T$) be the measurement scenario corresponding to the first $T-1$ 
($T$, respectively) measurements giving the outcome $0$. 
We compute the probabilities $p_{T-1}$ and $p_T$ that $\tau_{T-1}$ and $\tau_T$
are realized.  Then the probability that the one-way computation produces $0$
is precisely $p_T/p_{T-1}$. The computation is deterministic and takes $|V|^{O(1)}\exp[O(\cc(G))]$ time
by Lemma~\ref{res:singleQubitMeasure}.
\end{proof}

The main difference between the above lemma and Theorem~\ref{res:oneway}
is that the simulating complexity of the former is exponential in $\cc(G)$,
while that of the latter is exponential in $\tw(G)$. Since $\maxDegree(G)$ is 
not bounded in general, the lemma does not directly imply the theorem.
We shall reduce a one-way computation on a graph state $|G\rangle$
to a one-way computation on another graph state $|G'\rangle$, such
that $\maxDegree(G')=O(1)$ and $\tw(G')=O(\tw(G))$. Under this reduction,
the exponent in the simulating complexity is on the order of
$\cc(G')=O(\tw(G'))=O(\tw(G))$. 
Such a reduction was found in \cite{MarkovS:expansion}.
Let $G$ and $G'$ be two graphs. We call $G'$ an {\em expansion}
of $G$ if $G$ can be obtained from $G'$ by contracting
a set of edges that form a forest.

\begin{theorem}[\cite{MarkovS:expansion}]\label{res:expansion} Any undirected
simple graph $G=(V, E)$ has an expansion $G'=(V', E')$ 
such that $\maxDegree(G')\le 3$, $|V'|=O(|E|+|V|)$,
and $\tw(G')\le\tw(G)+1$.
Furthermore, such a graph $G'$ can be computed deterministically from $G$ in $(|V|)^{O(1)}\exp[O(\tw(G))]$ time.
\end{theorem}

In our application we need to insert a vertex 
into an edge that will be contracted during the transformation
of the graph $G'$ in the above theorem to $G$. This is to facilitate
the application of the following fact about graph states,
a proof for which is given in the Appendix.

\begin{proposition}[\cite{RB}]\label{res:measureInsert} 
Let $G$ be a graph obtained from a simple
undirected graph $G'$ by replacing a vertex $u\in V(G')$
with three vertices $v$, $w$, and $v'$, such that $w$ is adjacent
to $v$ and $v'$ only, and each vertex adjacent to $u$ in $G'$ 
becomes adjacent to either $v$ or $v'$, but not both.
Then $|G'\rangle$ can be obtained from $|G\rangle$ by an oblivious one-way computation
that makes $2$ measurements.
\end{proposition}

\begin{figure}[tbph]
\centering
\includegraphics[height=2.5in]{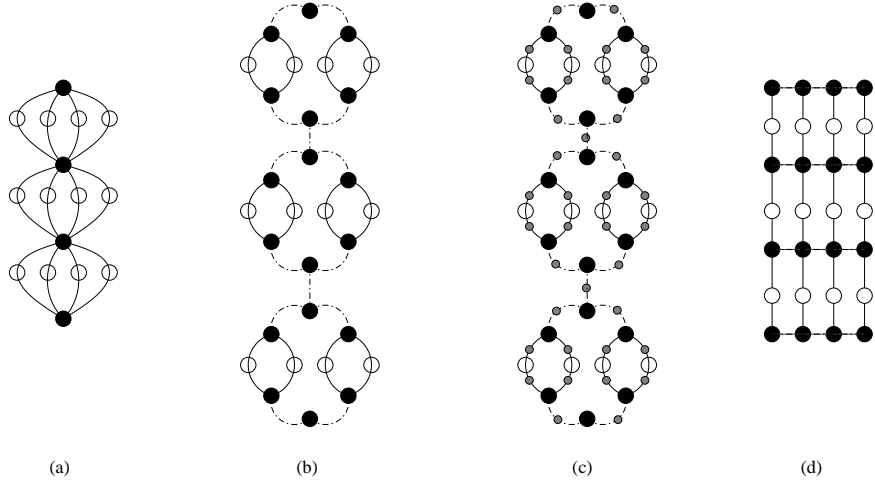}
\caption{To a graph with high-degree vertices in (a) we
apply the construction from \cite{MarkovS:expansion} to produce
a small-degree expansion in (b) that preserves treewidth. The graph
in (c) is obtained from (b) by inserting a vertex at each edge.
The corresponding graph state can lead to the graph state
of (a) through an oblivious one-way computation. The graph
in (d) illustrates that not every expansion of (a) preserves treewidth~\cite{MarkovS:expansion}.}
\label{fig:expansion}
\end{figure}

The use of expansion is illustrated in Figure~\ref{fig:expansion},
and summarized in the following Corollary.

\begin{corollary}\label{res:contractExpansion}
Let $G=(V, E)$ be a simple undirected graph.
There exists a graph $G_1=(V_1, E_1)$ such that
(a) $\maxDegree(G_1)\le 3$, (b) $|V_1|=O(|E|+|V|)$,
(c) $\tw(G_1)\le\tw(G)+1$,
(d) $G_1$ can be computed deterministically from $G$
in time $|V|^{O(1)}\exp[O(\tw(G))]$, and,
(e) $|G\rangle$ can be obtained by an oblivious one-way computation on $|G_1\rangle$,
\end{corollary}

\begin{proof} Let $G'=(V', E')$ be a graph
satisfying the properties in Theorem~\ref{res:expansion}.
Let $E'_1\subseteq E'$ be the set of edges contracting which
would transform $G'$ to $G$.
For each $e\in E'_1$, insert a vertex at $e$ (that is,
disconnect the end vertices of $e$ and connect them
to the new vertex). 

We show that the resulting graph $G_1$ satisfies the required properties.
Note that by Proposition~\ref{res:insert}, $\tw(G')=\tw(G_1)$.
Properties of $G'$ implies that Properties (a--e) hold.
The composition of the oblivious one-way computation
in Proposition~\ref{res:measureInsert} applied to the inserted
vertices transforms $|G_1\rangle$ to $|G\rangle$, and is itself
oblivious.
\end{proof}

We are now able to prove this section's main theorem, which restates Theorem~\ref{res:oneway}
and extends it to the case of oblivious one-way computation.
\begin{theorem}\label{res:onewayO} Let $G=(V, E)$ be a simple undirected graph. Then a one-way
computation on $G$ can be simulated by a randomized algorithm in time $|V|^{O(1)}\exp[O(\tw(G))]$.
The simulation can be made deterministic if the one-way computation is oblivious.
\end{theorem}
\begin{proof}
Let $G_1=(V_1, E_1)$ be a graph satisfying the properties stated in Corollary~\ref{res:contractExpansion}.
Thus $|G\rangle$ can be obtained from $|G_1\rangle$
through an oblivious one-way computation.
Therefore, the given one-way computation $P$ on $|G\rangle$ can be carried out by a one-way computation $P'$
on $|G_1\rangle$ which first produces $|G\rangle$ then continues executing $P$.
Note that $P'$ is oblivious if $P$ is.
By Lemma~\ref{res:onewaydegree}, $P'$ can be simulated by a randomized, or deterministic if
$P$ is oblivious, algorithm
in time $O(|V_1|^{O(1)}\exp[O(\cc(G_1))])$. 
Note that $\cc(G_1)=O(\tw(G_1))$, by Lemma~\ref{res:equivalence}, since $\maxDegree(G_1)\le 3$. 
Thus $\cc(G_1)=O(\tw(G))$, since $\tw(G_1)\le\tw(G)+1$.
Since $|V_1|=O(|V|+|E|)=O(|V|^2)$, the simulation time complexity is
$O(|V|^{O(1)} \exp[O(\tw(G))])$. 
\end{proof}

\section{Discussion}
  In this work we studied quantum circuits regardless
  of the types of gates they use, but with a focus on
  how the gates are connected. We have shown that 
  quantum circuits that look too similar to trees 
  do not offer significant advantage over classical
  computation. More generally, when solving a difficult
  classical problem on a quantum computer, one encounters
  an {\em inherent} trade-off between the treewidth and 
  the size of quantum circuits for solving it --- the smaller
  the quantum circuit, the more topologically sophisticated
  it must be. Investigating such trade-offs for specific problems 
  of interest is an entirely open and very attractive avenue for
  future research. Similar considerations may apply to classical circuits.
We conjecture that there are simple functions, such as modular exponentiation,
whose circuit realizations require large treewidth.

Furthermore, our work raises an intriguing possibility that the treewidth of
  some quantum circuits may be systematically reduced by restructuring
the circuit,
  while preserving the final result of the entire computation. 
  Perhaps, future research in this direction can
  clarify the limits to efficient quantum computation, while the
  tools developed in this context will be useful for practical tasks.

The pre-print of this paper~\cite{MarkovS05}
has lead to several follow-up results.
Jozsa~\cite{Jozsa06} and Aharonov et al.~\cite{Aharonov06}
gave alternative proofs for some of our theorems.
Furthermore, Aharonov et al.~\cite{Aharonov06},
and Yoran and Short~\cite{Yoran07} pointed out that
Quantum Fourier Transform (QFT) over $\mathbb{Z}_n$
admits approximate circuit realizations that, viewed as tensor
networks, have small treewidth.
Given the central role of QFT in known quantum algorithms, 
their results are somewhat unexpected and their implications
are yet to be fully explored. For example,
what type of circuits would remain efficiently simulatable
when interleaved with QFT circuits?
In general, as implied by Theorem~\ref{res:depth4} and~\ref{res:depth3},
the treewidth of a circuit may increase dramatically under composition.
Yoran and Short~\cite{Yoran07} have shown that this drawback may be avoided
in some cases.  Extending their result would deepen
our understanding of quantum speed-ups.

The important question of characterizing quantum states that are
universal (or efficiently simulatable) for one-way quantum computation
 remains unsolved.
In another follow-up thread, van den Nest et al.~\cite{denNestM06, denNestD06}
defined additional width-based parameters
of quantum states and demonstrated results for those parameters similar 
to Theorem~\ref{res:oneway}.
It is unlikely that the set of quantum states with small width-based parameters
includes all efficiently simulatable states
because a set of simulatable states of high widths
was identified recently by Bravyi and Raussendorf~\cite{Bravyi06}.
Nevertheless, it remains plausible that those width-based results
and their further extensions may be part of a classification theorem that gives
a complete characterization of efficiently simulatable states.

{\bf Acknowledgments.}
We thank George Viamontes and John Hayes for
motivating this study and helpful discussions.
We are grateful to Guifr{\'e} Vidal, Frank Verstraete, Ashwin Nayak, Tzu-Chieh Wei,
and the anonymous reviewers for many valuable comments, including
pointing out relevant previous works.
Y.~S. is grateful to Alexei Kitaev for introducing to him
the general concept of tensor networks, and to Peng-Jun Wan
for useful comments. I.~M. is grateful
to Ike Chuang for useful discussions.

\appendix
\section{Proof of Proposition~\ref{res:insert}}
Recall that a minor of a graph $G$ is a graph obtained from a subgraph of $G$
by contracting edges. A basic property of treewidth is that it does
not increase under taking minors~\cite{RSII}.

\begin{proofof}{Proposition~\ref{res:insert}}
Let $G'$ be the graph resulting from the contractions.
Since $G'$ is a minor of $G$, $\tw(G')\le\tw(G)$ (\cite{RSII}).
If $\tw(G')=1$, then $G'$ is a non-empty forest (otherwise $G$ has a triangle
minor, thus $\tw(G)\ge 2$). Thus $G$ is also
a non-empty forest and $\tw(G)=1=\tw(G')$.
Suppose $\tw(G')\ge 2$. Let $\mathcal{T}$ be a tree decomposition for $G'$.
We obtain a tree decomposition $\mathcal{T}'$
for $G$ by inserting a bag containing $\{u, w, v\}$,
and connecting it to a bag that contains $\{u, v\}$.
One can verify directly that the three conditions $(T_1-T_3)$
that define tree decompositions hold for $\mathcal{T}'$.
Since the width of $\mathcal{T}'$ is no more than that of $\mathcal{T}$,
we have $\tw(G)\le \tw(G')$. Therefore, $\tw(G)=\tw(G')$.
\end{proofof}

\section{Proof of Proposition~\ref{res:measureInsert}}
\begin{proofof}{Proposition~\ref{res:measureInsert}}
Denote by $\bar G$ the subgraph of $G'$ induced by $V(G')-\{u\}$.
Let $A$ and $A'$ be vertices in $V(G')-\{u\}$ that are adjacent
to $v$ and $v'$, respectively, in $G$. Note that $A\cap A'=\oslash$,
thus $A\oplus A'=A\cup A'$ is the neighborhood of $u$ in $G'$. Also,
$v\not\in A'$ and $v'\not\in A$.

Starting with $|G\rangle$, we first measure $\sigma_x$ on $w$.  If the outcome is $+1$,
the resulting state is
\begin{equation}\label{eqn:state1} |\phi_1\rangle\defeq |00\rangle_{v'v}|\bar G\rangle + 
|11\rangle_{v'v} \sigma_z[A\oplus A']|\bar G\rangle.
\end{equation}
Otherwise, the resulting state is
\[ |01\rangle_{v'v} \sigma_z[A]|\bar G\rangle + |10\rangle_{v'v}\sigma_z[A']|\bar G\rangle,\]
which can be brought to $|\phi_1\rangle$ by $\sigma_x[v]\sigma_z[A]$.
We then measure $\sigma_x[v']$ on $|\phi_1\rangle$.
If the outcome is $+1$, then the resulting state is precisely $|G'\rangle$. Otherwise it is
\[ |0\rangle_v |\bar G\rangle - |1\rangle \sigma_z[A\oplus A']|\bar G\rangle,\]
which can be brought to $|G'\rangle$ by $\sigma_z[v]$.
The four outcomes of the two measurements have equal probability ($1/4$).
Thus the one-way computation is oblivious.
\end{proofof}

\begin{thebibliography}{10}

\bibitem{AaronsonG04}
S.~Aaronson and D.~Gottesman.
\newblock Improved simulation of stabilizer circuits.
\newblock {\em Physical Review A}, 70:052328, 2004.

\bibitem{AKN98a}
D.~Aharonov, A.~Kitaev, and N.~Nisan.
\newblock Quantum circuits with mixed states.
\newblock In {\em Proceedings of the 31th Annual ACM Symposium on the Theory of
  Computation (STOC)}, pages 20--30, 1998.

\bibitem{Aharonov06}
D.~Aharonov, Z.~Landau, and J.~Makowsky.
\newblock The quantum FFT can be classically simulated.
\newblock Preprint: quant-ph/0611156.

\bibitem{AlberB+02}
J.~Alber, H.~L. Bodlaender, H.~Fernau, T.~Kloks, and R.~Niedermeier.
\newblock Fixed parameter algorithms for dominating set and related problems on
  planar graphs.
\newblock {\em Algorithmica}, 33(4):461--493, 2002.

\bibitem{Arnborg85}
S.~Arnborg.
\newblock Efficient algorithms for combinatorial problems on graphs with bounded decomposability --- a survey.
\newblock {\em BIT}, 25(1):2--23, 1985.

\bibitem{ArnborgCP87}
S.~Arnborg, D.~G. Corneil, and A.~Proskurowski.
\newblock Complexity of finding embeddings in a {$k$}-tree.
\newblock {\em SIAM Journal on Algebraic and Discrete Methods}, 8(2):277--284, 1987.

\bibitem{BR}
H.J.~Briegel and R.~Raussendorf.
\newblock Persistent entanglement in arrays of interacting particles.
\newblock {\em Physical Review Letters}, 86, 910--913, 2001.

\bibitem{Bodlaender06}
H.~L. Bodlaender.
\newblock Treewidth: characterizations, applications, and computations.
\newblock Technical Report UU-CS-2006-041, Universiteit Utrecht.

\bibitem{Bravyi06}
S.~Bravyi and R.~Raussendorf.
\newblock On measurement-based quantum computation with the toric code states.
\newblock Preprint: quant-ph/0610102. 

\bibitem{BroeringL03}
E.~Broering and S.~Lokam.
\newblock Width-based algorithms for {SAT} and {Circuit}-{SAT} (extended
  abstract).
\newblock In {\em Sixth International Conference on Theory and Applications of
  Satisfiability Testing (SAT 2003),
Springer-Verlag Lecture Notes in Computer Science
(LNCS)}, volume~2919, pp. 162--171, 2004.

\bibitem{Dechter99}
R.~Dechter.
\newblock Bucket elimination: a unifying framework for reasoning.
\newblock {\em Artificial Intelligence}, 113(1-2):41--85, 1999.

\bibitem{DuanR:2005}
L.-M.~Duan and R.~Raussendorf.
\newblock Efficient quantum computation with probabilistic quantum gates.
\newblock {\em Physical Review Letters}, 95:080503, 2005.

\bibitem{Gottesman:1998:simulate}
D.~Gottesman.
\newblock The {H}eisenberg representation of quantum computers.
\newblock In S.~P. Corney, R.~Delbourgo, and P.~D. Jarvis, editors, {\em
  Group22: Proceedings of the XXII International Colloquium on Group
  Theoretical Methods in Physics}, pages 32--43, Cambridge, MA, 1999.
  International Press.
\newblock Long version: quant-ph/9807006.

\bibitem{FennerGHZ04}
S.~F.~F. Green, S.~Homer, and Y.~Zhang.
\newblock Bounds on the power of constant-depth quantum circuits.
\newblock Preprint: quant-ph/0312209, 2004.

\bibitem{Joshi95}
A.~W. Joshi.
\newblock {\em Matrices and tensors in physics}.
\newblock Halsted Press [John Wiley \& Sons], New York-London-Sydney, 1975.

\bibitem{Jozsa06}
R.~Jozsa.
\newblock On the simulation of quantum circuits.
\newblock Preprint: quant-ph/0603163.

\bibitem{JL02a}
R.~Jozsa and N.~Linden.
\newblock On the role of entanglement in quantum computational speed-up.
\newblock {\em Proceedings of the Royal Society of London, Series A}, 459: 2011-2032, 2003.

\bibitem{KrishnaswamyVMH05}
S.~Krishnaswamy, G.~F.~Viamontes, I.~L.~Markov and J.~P.~Hayes.
\newblock Accurate reliability evaluation and enhancement via probabilistic
 transfer matrices.
\newblock {\em Proc. Design Automation and Test
in Europe (DATE)}, pp. 282-287, Munich, Germany, March 2005.

\bibitem{LubotzkyPS88}
A.~Lubotzky, R.~Phillips, and P.~Sarnak.
\newblock Ramanujan graphs.
\newblock {\em Combinatorica}, 8(3):261--277, 1988.


\bibitem{MarkovS05}
I.~L.~Markov and Y.~Shi.
\newblock Simulating quantum computation by contracting tensor networks.
\newblock Pre-print: quant-ph/0511069.

\bibitem{MarkovS:expansion}
I.~L.~Markov and Y.~Shi.
\newblock Constant degree graph expansions and treewidth.
\newblock Manuscript.

\bibitem{denNestD06}
M.~van~den~Nest, W.~D{\"u}r, G~ Vidal,and  H~ J.~ Briegel.
\newblock Classical simulation versus universality in measurement based quantum computation.
\newblock {\em Physical Review A}, 75:012337, 2007.

\bibitem{denNestM06}
M.~van~den~Nest, A.~Miyake, W.~D{\"u}r and H.~J~ Briegel. 
\newblock Universal resources for measurement-based quantum computation.
\newblock {\em Physical Review Letters}, 97:150504, 2006.

\bibitem{NielsenC}
M.~A.~Nielsen and I.~L.~Chuang.
\newblock {\em Quantum Computation and Quantum Information}.
\newblock Cambridge University Press, Cambridge, England, 2000.

\bibitem{Porras05}
\newblock D.~Porras, F.~Verstraete and J.~I.~Cirac.
\newblock Renormalization algorithm for the calculation of spectra of interacting quantum systems.
\newblock Preprint: cond-mat/0504717.

\bibitem{RB}
R.~Raussendorf and H.~J.~Briegel.
\newblock A one-way quantum computer.
\newblock {\em Physical Review Letters}, 86, 5188--5191, 2001.

\bibitem{RSII}
N.~Robertson and P.~D. Seymour.
\newblock Graph minors. {II}. Algorithmic aspects of tree-width.
\newblock {\em Journal of Algorithms}, 7(3):309--322, 1986.

\bibitem{RSIII}
N.~Robertson and P.~D. Seymour.
\newblock Graph minors. {III}. {P}lanar tree-width.
\newblock {\em Journal of Combinatorial Theory, Series B}, 36(1):49--64, 1984.

\bibitem{RSX}
N.~Robertson and P.~D. Seymour.
\newblock Graph minors. {X}. {O}bstructions to tree-decomposition.
\newblock {\em Journal of Combinatorial Theory, Series B}, 52(2):153--190, 1991.

\bibitem{Rose70}
D.~J. Rose.
\newblock Triangulated graphs and the elimination process.
\newblock {\em Journal of Mathematical Analysis and Applications}, 32:597--609, 1970.

\bibitem{RoychowdhuryV}
V.~P. Roychowdhury and F. Vatan.
\newblock Quantum formulas: a lower bound and simulation.
\newblock {\em SIAM Journal on Computing}, 31(2): 460--476, 2001.

\bibitem{Terhal:2002:simulate}
B.~M. Terhal and D.~P. DiVincenzo.
\newblock Classical simulation of noninteracting-fermion quantum circuits.
\newblock {\em Physical Review A}, 65:32325--32334, 2002.

\bibitem{Terhal:2002:constant}
B.~M. Terhal and D.~P. DiVincenzo.
\newblock Adaptive quantum computation, constant depth quantum circuits and
  {A}rthur-{M}erlin games.
\newblock {\em Quantum Information and Computation}, 4(2):134--145, 2004.

\bibitem{Valiant:2002:simulate}
L.~G. Valiant.
\newblock Quantum circuits that can be simulated classically in polynomial
  time.
\newblock {\em SIAM Journal on Computing}, 31(4):1229--1254, Aug. 2002.

\bibitem{Verstraete04}
\newblock F.~Verstraete and J.~I.~Cirac.
\newblock Renormalization algorithms for Quantum-Many Body Systems in two and higher dimensions.
\newblock Preprint: cond-mat/0407066.

\bibitem{VerstraeteG04}
\newblock F.~Verstraete, J.~J.~Garcia-Ripoll and J.~I.~Cirac.
\newblock Matrix product density operators: simulation of 
finite-temperature and dissipative systems.
\newblock {\em Physical Review Letters}, 93:207204, 2004.

\bibitem{Vidal03}
G.~Vidal.
\newblock Efficient classical simulation of slightly entangled quantum computations.
\newblock {\em Physical Review Letters}, 91:147902, 2003.

\bibitem{Vidal04}
G.~Vidal.
\newblock Efficient simulation of one-dimensional quantum
many-body systems.
\newblock {\em Physical Review Letters}, 93:040502, 2004.

\bibitem{Walther05}
P.~Walther, K.J.~Resch, T.~Rudolph, E.~Schenck, H.~Weinfurter,
V.~Vedral, M.~Aspelmeyer, and A.~Zeilinger.
\newblock Experimental one-way quantum computing.
\newblock {\em Nature}, 434, 169--176, 2005.

\bibitem{Yoran07}
N.~Yoran and A.~Short.
\newblock Efficient classical simulation of the approximate quantum Fourier transform.
\newblock {\em Physical Review A}, 76:042321, 2007.

\bibitem{Yao93}
A.~Yao.
\newblock Quantum circuit complexity.
\newblock {\em Proceedings of the 34th Annual Symposium on
                 Foundations of Computer Science}, 352--361, 1993.

\bibitem{Zwolak04}
M.~Zwolak and G.~Vidal.
\newblock Mixed-State dynamics in one-dimensional quantum lattice
systems: a time-dependent superoperator renormalization algorithm.
\newblock {\em Physical Review Letters}, 93:207205, 2004.
\end{thebibliography}
\end{document}